\documentclass{article}

\title{Evolution of Digital Advertising Strategies\\during the 2020 US Presidential Primary}

\author{NaLette Brodnax$^1$ \and Piotr Sapiezynski$^2$}
\date{\small{$^1$McCourt School of Public Policy, Georgetown University \\
      $^2$Khoury College of Computer Sciences, Northeastern University}}

\usepackage[utf8]{inputenc}
\usepackage{amsmath}
\usepackage{graphicx}
\usepackage{rotating}
\usepackage{amsmath}
\usepackage{booktabs}
\usepackage{url}
\usepackage{titling}
\usepackage[style=authoryear]{biblatex}

\begin{document}
\maketitle

\begin{abstract}
Political advertising on digital platforms has grown dramatically in recent years as campaigns embrace new ways of 
%CUT precisely 
targeting supporters and potential voters. Previous scholarship shows that digital advertising has both positive effects on democratic politics through increased voter knowledge and participation, and negative effects through user manipulation, opinion echo-chambers, and diminished privacy. However, research on election campaign strategies has focused primarily on traditional media, such as television.
%CUT advertisements
Here, we examine how political campaign dynamics have evolved in response to the growth of digital media by analyzing the advertising strategies of US presidential election campaigns during the 2020 primary cycle. To identify geographic and temporal trends, we employ regression analyses of campaign spending across nearly 600,000 advertisements published on
%CUT the 
Facebook.
%CUT platform. 
We show that campaigns heavily target voters in candidates’ home states during the ``invisible primary'' stage before shifting to states with early primaries.
\end{abstract}

\section{Introduction}
In recent years, presidential campaigns in the United States have shifted from traditional forms of advertising---such as television, radio, and print---to digital media. Historically, campaigns have spent the largest proportion of their advertising budgets on television commercials, as they are most effective in reaching a large number of voters. However, the proportion of ad spending devoted to television has declined in each election year since 2008~\parencite{cassino2017a}.  Over the same period, the proportion spent on digital advertising increased from 0.2\% to 14.4\% in 2016 and is projected to account for 28\% of political ad spending in 2020.% (see Figure~\ref{fig:budget_fracs}).  

Digital advertising has a number of advantages over television and other traditional media. 
Most importantly, it allows campaigns to precisely target voters using a range of tools made available to them by ad platforms.
Campaigns can choose the target audience based on their location, %CUT (city, region, state, nationwide), 
age, and gender, as well as a number of interest based parameters (such as inferred political alignment). 
%CUT Further, campaigns can target particular users based on their personally identifiable information (PII), such as email addresses, phone numbers or names, as well as promote ads to other users similar to those whose PIIs they've obtained~\parencite{martinez2018}. 
%CUT Social media platforms allow users to interact with ads by clicking, commenting, or sharing. As a result, campaigns receive immediate feedback on an ad's ability to engage potential voters and can use it more efficiently allocate resources~\parencite{martinez2018, erdody2018a}. 
By clicking, commenting, or sharing ads, social media users provide campaigns with immediate feedback on ads' ability to engage potential voters, and the campaigns, in turn, can use it more efficiently allocate resources~\parencite{martinez2018, erdody2018a}. 
Finally, campaigns can advertise on digital platforms with relatively small budgets, in contrast to television advertising where budgets can run from hundreds of thousands to millions of dollars. 
%CUT Digital ads' lower relative cost has been critical for lesser-known candidates, as an increased web presence corresponds to increased financial and electoral support, serving as an equalizing force for long-shot candidates~\cite{christenson2014a, paolino2003a}. %\nalette{Not happy with this paragraph wording.}
Due to their lower cost, digital ads have served as an equalizing force for long-shot candidates~\parencite{christenson2014a, paolino2003a}.

Despite dramatic growth in digital advertising, scholars know little about how political campaigns leverage its unique features. To address this gap, we consider two questions. First, we 
%CUT consider 
investigate whether campaigns utilize the option to target voters by location in a manner consistent with traditional forms of advertising.
%CUT, such as television commercials.
Second, we examine whether campaigns’ advertising strategies shift over time as they consider the next immediate need.
%CUT or receive feedback on the effectiveness of their ads.
To answer these questions, we examine Facebook advertising by U.S. presidential campaigns during the 2020 primary election cycle. Facebook accounts for the largest share of digital advertising due its ease of use and the size of its user base~\parencite{erdody2018a}.  
Using the Facebook Political Advertising Library ~\parencite{FacebookAdLibrary}, we observe nearly 600,000 advertisements published by 26 presidential primary campaigns from January 1, 2019 through ``Super Tuesday,'' March 3, 2020. 
%CUT The library reports distribution via a metric known as an \textit{impression}. For a given ad, the impression count refers to the number of times that the ad appeared in Facebook users' feeds. 
For each ad, the library reports the estimated number of impressions---the number of times that the ad appeared in users' feeds.
In addition to estimated impressions,
%CUT the estimated number of impressions, 
the library reports the approximate cost, as well as the locations and demographic characteristics, such as age and gender, of users who viewed each ad.

Research on campaign advertising strategies has utilized a range of data sources, the most 
prominent being 
% CUT prominently 
television advertising archives~\parencite{goldstein2011a,fowler2017a}.
%CUT hosted by the University of Wisconsin ~\parencite{goldstein2011a} and Wesleyan University~\parencite{fowler2017a}. 
These archives collect and encode data on television ads
%CUT that aired on a daily basis 
across 210 designated market areas (DMAs), including location, cost, and content. Since some DMAs span multiple states, this data has limited utility for analyzing advertising strategies by state. Because states award both primary delegates and general electors, state-level outcomes are important for answering questions about presidential campaign strategies. The Facebook Ad Library provides the same level of detail but also allows us to compare state-level advertising by each campaign over time.

We find that campaigns target voters by state and that their strategy shifts over time. Early in the primary season, campaigns spend a larger share of their budget in the candidate's home state where their fundraising efforts are boosted by name recognition. As the first wave of state caucuses and primary elections approach, campaigns shift funding to states with early primaries and, to a lesser extent, swing states. 

\section{Political Advertising Strategies}\label{sec:background}

How do presidential candidates win primaries? This question has been the subject of considerable debate since the modern era of presidential nominations began in 1972. Some argue that campaigns use media network coverage, fundraising, and endorsements to court party elites, who then coordinate to informally select a nominee prior to the first caucuses and primaries ~\parencite{aldrich2009a,dowdle2009a}. However, in response to a series of campaign finance reforms, this party-centered ``invisible primary’’ has become more candidate-centered, with campaigns courting not only party elites but also interest groups and individual donors ~\parencite{la-raja2015a}.  Candidates also seek to exploit the timing of primaries---particularly the early races in Iowa and New Hampshire---to generate momentum ~\parencite{bartels1988a}. We consider how campaigns use advertising in the midst of these competing dynamics to pursue a set of complementary and sequential goals: establishing viability during the invisible primary, generating momentum during the early primaries, and outraising their opponents. We posit that campaigns use explicit geographic and temporal strategies to achieve these ends.

\subsection{Signaling Viability}
Given evidence that campaigns with greater web presence receive more contributions, campaigns see digital advertising as a key fundraising avenue ~\parencite{christenson2014a}. Throughout the primary cycle, candidates focus 
%CUT intently 
on raising money, as the amounts
%CUT of money 
raised and in cash reserves have been reliable predictors of winning the primary election
%CUT in the past
~\parencite{adkins2001a}.  In addition to supporting campaign expenditures, fundraising sends important signals about candidate viability.   Candidates must report their fundraising totals to the Federal Election Commission (FEC) on a monthly basis, allowing the public to gauge whether candidates have financial support from a wide array of contributors.  Further, candidates must 
%CUT raise a certain amount of money 
meet certain fundraising thresholds
to be formally recognized by their political party. During the 2020 primary season, the Democratic National Committee required candidates to meet specific fundraising and polling thresholds in order to participate in televised debates ~\parencite{scherer2019a}.

Previous research suggests that campaigns target specific geographic regions for fundraising~\parencite{gimpel2006a, cho2007a, lin2017a}. 
We posit that campaigns will concentrate their digital advertising efforts where they are most well known and therefore contributions are most likely---the candidate's home state---during the invisible primary. 
We expect that this effect will be most pronounced prior to the first debate, where outsider candidates would be best able to educate the public about their candidacy. 
\textit{Hypothesis 1: Campaigns target contributors in their home states.}
We also consider whether this geographic strategy has a temporal component. \textit{Hypothesis 2: Campaigns’ geographic targeting shifts over time.} 
%CUT Lesser known candidates must first generate awareness among the general public before seeking contributions. 

\subsection{Gaining Momentum}
Advertising during the earliest primaries is a key strategy for established
%CUT , well-known 
candidates to eliminate competition, or for lesser-known candidates looking to make a name for themselves ~\parencite{paolino2003a}. Although voting is spread out over five months, delegate allocation is front-loaded. 
%CUT For the 2020 Democratic presidential primary season,
Approximately half of all delegates were allocated during the first third of the 2020 Democratic presidential primary season ~\parencite{almukhtar2019a}. As a result, campaigns may concentrate their advertising spending in states with earlier primaries in order to establish momentum. A study of primaries conducted between 1980 and 1996 shows that the New Hampshire primary, which takes place
%CUT very 
early in the primary season, is correlated with the ordinal ranking of candidates ~\parencite{adkins2001a}. In addition, the nomination of Barack Obama in 2008 has been attributed in part to his campaign's use of digital advertising to establish early momentum ~\parencite{aldrich2009a}. \textit{Hypothesis 3: Campaigns target voters in states with early primaries.}

During the general election, most states award all of their electoral votes to the winner, giving presidential candidates an incentive to campaign in states with large numbers of electors. Campaigns also allocate their resources based on whether a state is considered a battleground or part of the party base ~\parencite{shaw1999a, shaw2006a}. However, campaign strategies during the primary may or may not follow the logic of the general election. Since campaigns are not competing against another party, battleground or ``swing'' states may warrant fewer resources than states with large numbers of delegates. Further, a recent study shows that campaigning in uncontested states can lead to increased campaign contributions ~\parencite{urban2014a}. On the other hand, campaigns may begin advertising in swing states in order to establish momentum going into the general election. \textit{Hypothesis 4: Campaigns target voters in swing states.}

\section{Measures}\label{sec:methods}

Consider a campaign $i$ with a monthly budget to allocate to different states $\{1, ..., S\}$. The number of observations for each month is equal to $N_i\times S$, where $N_i$ is the number of active campaigns $i$ and $S=51$, the number of US states plus the District of Columbia.

\subsection{Dependent variable}
Let $Y_{is}$ be the population-normalized proportion of the advertising budget allocated to state $s$ by campaign $i$. If a campaign spends money in direct proportion to the number of inhabitants, then $Y_{is}=1$. 
%CUT When campaigns spend more or less than expected given a state's population, then $Y_{is}\neq 1$.
We use the budget proportion as our dependent variable rather than the dollar amount because it allows us to compare candidates whose budgets vary by orders of magnitude.
We normalize the proportion by each state's population to compare how much value campaigns assign to voters living in states whose populations vary by orders of magnitude. See Supplementary Information Section S2 for the data description and Section S3 for additional discussion of the dependent variable construction.

\subsection{Independent variables}
Each state $s$ has a set of characteristics taken into consideration by a campaign.  The variable \textit{Home} is a binary indicator with a value of $1$ if $s$ is the home state for the campaign $i$. \textit{Swing} indicates whether state $s$ is considered a swing state, meaning it is not characterized by a clear preference towards Democratic or Republican candidates in recent elections. We also include \textit{February primary}  and \textit{Super Tuesday primary} to indicate whether state $s$ holds an early caucus or primary.  

\subsection{Model specification}\label{sec:model}
To estimate the budget proportion we fit the following using ordinary least squares:
\begin{align*}
	Y_{is} &= \alpha + \beta_1 Home_s + \beta_2 Swing_s\\ &+ \beta_3 February_s + \beta_4 Super Tuesday_s + \epsilon_{is}
\end{align*}

The intercept is the ratio of state budget to state population. Not accounting for state characteristics, this value should be approximately one, reflecting our expectation that campaigns will spend more money in states with larger populations. 
The coefficients $\{\beta_1, \beta_2, \beta_3, \beta_4\}$ are our quantities of interest. A coefficient greater than zero means that campaigns spent more to target the same share of residents in states with that characteristic. 

\section{Results}\label{sec:results}
We report 
%CUT the model 
results for the aggregate of all data in Table~\ref{table:overall}. We then report results from a series of independent monthly and weekly models to reflect the temporal changes.

\subsection{Aggregate model}
%CUT We report our regression estimates in Table~\ref{table:overall}. 
If ad spending were proportional to the population, we would expect the \textit{Intercept} value to be near 1 and for the coefficients of the other variables to be 0.
%CUT We observe that 
Shown in Table~\ref{table:overall}, the \textit{Intercept} is slightly smaller than 1, suggesting that on average, states 
%CUT which are not home to a candidate, are not considered ``swing'', and have relatively late primaries 
see somewhat less ad spending than expected given their populations.

\begin{table}[!ht] 
\centering 
  \caption{Relative proportion of advertising budget} 
  \label{table:overall} 
\begin{tabular}{@{\extracolsep{5pt}}lc } 
\midrule
 Home & 6.408*** \\ 
  & (0.487) \\ 
  Swing & 0.325** \\ 
  & (0.151) \\ 
  February Primary & 3.293*** \\ 
  & (0.267) \\ 
  Super Tuesday Primary & --0.293** \\ 
  & (0.147) \\ 
  Intercept & 0.813*** \\ 
  & (0.085) \\ 
 \hline \\[-1.8ex] 
Observations & 1,482\\
\bottomrule
%\addtabletext{$^{*}$p$<$0.1; $^{**}$p$<$0.05; $^{***}$p$<$0.01}
$^{*}$p$<$0.1; $^{**}$p$<$0.05; $^{***}$p$<$0.01
\end{tabular} 
\end{table} 

\subsubsection{Campaigns spend considerably more in home states.}
Given the large magnitude and statistical significance of the \textit{Home} coefficient, we conclude that campaigns spend significantly more on ads in the candidates' home states, in line with our hypothesis (\textit{H1}).
A hypothetical candidate whose home state is not a swing state and is late in the primary calendar would spend a fraction of their budget per capita that is 7.2 times higher than in another non-swing, late-primary state ($6.408+0.813\approx7.2$).

\subsubsection{Earliest primaries more consequential than Super Tuesday.} Our third hypothesis posited that campaigns target voters in states with early primaries. We considered two types of calendar effects: (1) targeting in the first few primary states where the number of delegates is small but 
signaling candidate viability is important; and (2) targeting in states that can 
%CUT further 
establish momentum by allocating a large number of delegates relatively early in the cycle.  The positive coefficient for \textit{February Primary} and negative coefficient for \textit{Super Tuesday Primary} suggest that campaigns embraced the first strategy---targeting the first few primaries, particularly in Iowa, New Hampshire, Nevada, and South Carolina. Campaigns allocated roughly four times the per capita budget to target voters in these states ($3.293+0.813\approx4.1$). In contrast, campaigns spent slightly less than average ($\beta_3=-0.293$) per capita in Super Tuesday states. States with February primaries were targeted heavily, but still less so than candidates' home states. 
\subsubsection{Campaigns target swing states, but less so than home states.}
The positive and statistically significant value of the \textit{Swing state} coefficient supports our fourth hypothesis, that candidates target voters in swing states
%CUT specifically 
in a manner disproportionate to their populations. However, the low magnitude of the coefficient ($\beta_2=0.325$) 
%CUT is just half of a tenth the size of the \textit{Home} coefficient, suggesting 
suggests that battleground status alone is not a strong predictor of ad spending during the primary cycle. This may be explained by the fact that swing states are
%CUT much 
more critical for the general election than the primary election. Swing states may
%CUT therefore 
warrant more digital advertising during the general election, after the major party nominees have been determined.

\subsection{Temporal model}
\begin{figure*}[!t]
	\centering
	\includegraphics[width=1\linewidth]{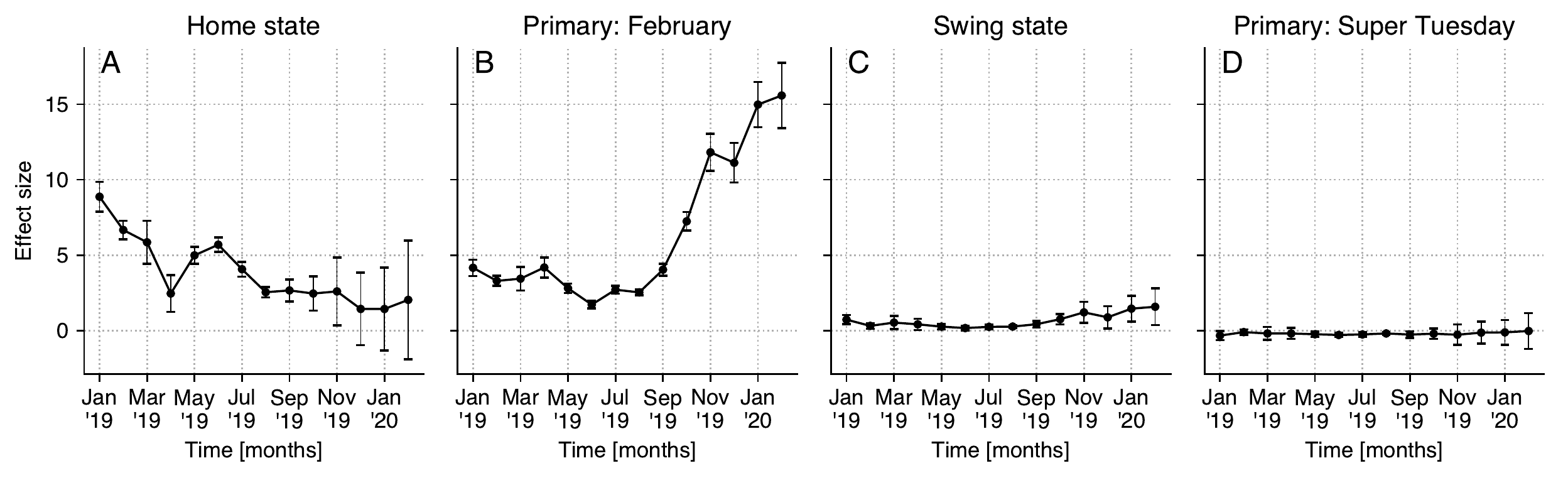}	
	\caption{In January, campaigns spent as much as 9 times more in their home states than expected given their population, but the importance of home states drops over time. In contrast, states with primaries in February see increasing spends over time, as do, to a lesser extent, the swing states. }
	\label{fig:temporal_model}
\end{figure*}

To examine our second hypothesis, that campaigns' targeting strategies shift over time, we fit the 
%aggregate 
model 
%specification 
to each month of data independently. Our results provide 
%strong 
evidence that campaigns alter their strategies over time---heavily targeting home states early in the primary cycle and shifting to early primary states and swing states as the first primaries and caucuses approach in February. Further, we show that advertising in Super Tuesday states remains flat throughout the duration of the time period we study. Figure~\ref{fig:temporal_model} presents the results of this analysis, where the panels reflect monthly coefficients and 95\% confidence intervals. 

\subsubsection{Campaigns target home states during first phase of calendar.}
It is instructive to look at the 14-month period in two phases: the first from January through July 2019, and the second from August 2019 through February 2020. Phase~I is characterized by heavy to moderate spending in home states and moderate spending in February primary states. During this phase, spending in swing states (other than February primary states) and Super Tuesday primary states remained constant, with coefficients at or near zero over the 7-month period. Panel A of Figure~\ref{fig:temporal_model} shows that political advertising is the most intense in the home state of each candidate in the beginning of 2019: more than a year before the first race, campaigns spent 9 times the average budget proportion per capita in these states. 

The overall pattern in the \textit{Home} coefficient during Phase~I suggests that campaigns 
%CUT may have 
utilized digital advertising as a strategy to qualify for the Democratic debates. Following the large spending boost in January 2019, we observe a decline followed by a rise in May 2019 and a second peak in June 2019.  This second peak corresponds to a push by campaigns to meet polling and fundraising thresholds for the first debates, which were held on June 26 and 27, 2019. 
%CUT According to guidelines set by 
The Democratic National Committee (DNC) required that candidates, ``register 1\% or more support in three polls released between January 1, 2019, and 14 days prior'' to the first debate ~\parencite{dnc2019a}. 
%CUT This establishes an initial qualification deadline June 12, 2019. 
Candidates also required
%CUT must also generate enough grassroots fundraising to acquire 
``(1) 65,000 unique donors; and (2) a minimum of 200 unique donors per state in at least 20 U.S. states.'' Since the unique donor threshold is much higher than the individual state donor minimum, campaigns may have targeted their home states to attract the 
%CUT roughly 60,000
number of donors needed. 
%CUT beyond the individual state minimums. 

In addition to debate qualification thresholds, campaigns' incentives were impacted by a historically large field of major candidates. The DNC set a cap of 20 candidates for the first set of debates.
%CUT ~\parencite{dnc2019a}. 
Since there were 26 major candidates, campaigns would have sought to meet and even exceed the qualification thresholds to ensure debate participation. Failure to meet debate thresholds may explain why, after the first debate, a substantial number of candidates
%CUT reduced or eliminated the number of published 
stopped publishing advertisements on Facebook.
%CUT advertisements on the Facebook platform.% (see Figure \ref{fig:candidate_count}).

\begin{figure}[!ht]
	\centering
	\includegraphics[width=.7\linewidth]{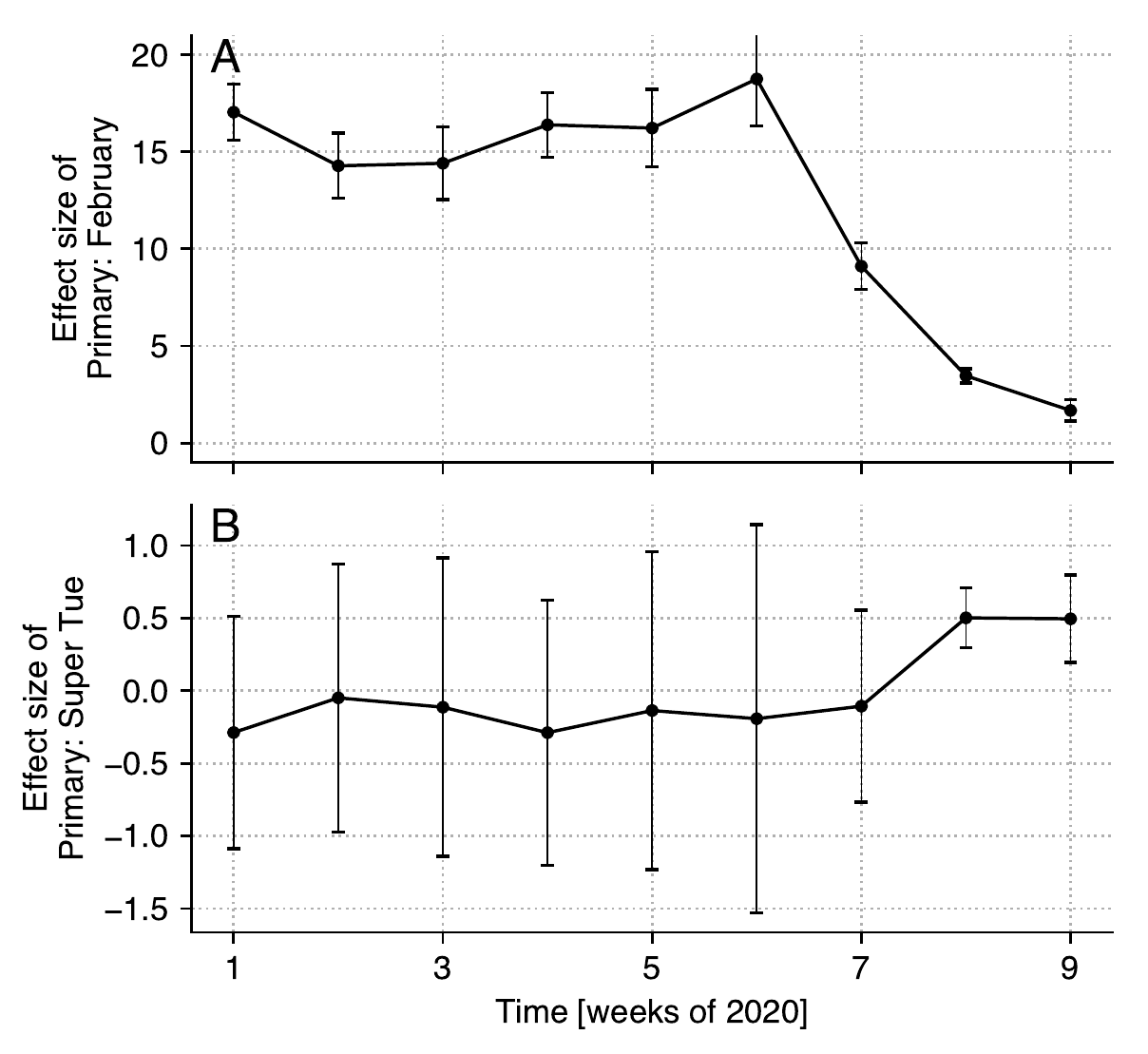}	
	\caption{A) The effect size of a primary in February is high through the first weeks of 2020. After each actual primary event, the effect size drops and nearly disappears by the last week of February/first week of March. B) On the other hand, the states that hold their primaries on Super Tuesday only see increased advertising in the last two weeks leading up to the date.}
	\label{fig:weekly_temporal_model}
\end{figure}

\subsubsection{February primary states targeted heavily in second phase.}
During Phase~II of the cycle, between August 2019 and March 2020, campaigns spent heavily in February primary states (Figure~\ref{fig:temporal_model}B), a majority of which are also swing states, and spent slightly more in other swing states (Figure~\ref{fig:temporal_model}C). Home state spending remained constant at a quarter of its Phase~I peak, and spending in Super Tuesday states remained constant with coefficients at or near zero. By the end of Phase~II, campaigns spent up to 16 times the proportionate baseline in states with the earliest primaries. This is three times more advertising than the February primary state baseline in June 2019, the peak at the end of Phase I.

To examine these effects more closely, we focus on the weeks leading up to the first primaries in February and March. Figure~\ref{fig:weekly_temporal_model}A shows that the states with a February primary receive 15 to 18 times more spending than expected given their populations, up to the first week of February.
After the Iowa caucuses, the focus switches to the remaining February states and as campaigns stop showing ads in Iowa, the effect size nearly disappears by the end of February.
In contrast, Figure~\ref{fig:weekly_temporal_model}B shows that the Super Tuesday states do not receive significantly more spending until the last week of February and first week of March, indicating that the campaigns focus spending on locations that are immediately next in the primary calendar.

\section{Discussion}\label{sec:analysis}

Our analysis demonstrates that presidential primary campaigns employ a range of digital advertising strategies and shift these strategies over time. Consistent with earlier research, we find that campaigns use digital media to target voters in states with early primaries, and to a lesser extent, voters in swing states. In contrast to earlier studies, we show that campaigns not only heavily target voters in candidates' home states, but do so early in the primary cycle. 

Although digital media spending and television spending are difficult to compare directly, an exploratory analysis of television advertising data suggests that early home-state spending is a relatively new phenomenon. For comparison, we examined estimated spending on television advertisements during the 2008 and 2012 presidential primary cycle ~\parencite{goldstein2011a,fowler2017a}. 
%CUT Restricting our sample to major candidates (those spending at least \$100K during the first 14 months of the primary season) and ads purchased by campaigns rather than parties or committees, we find that campaigns did not use television advertisements to target voters in candidates' home states. 
We find that the campaigns of major candidates did not use television advertisements to target voters in candidates' home states.
Additional details about the comparison are provided in the Supplementary Information Section S4. 
Recent research comparing campaigns' use of both television and Facebook advertising finds that campaigns use digital advertising earlier and use ad content to promote the candidate rather than attack opponents or promote issues ~\parencite{fowler2020}. These findings are consistent with our perspective that early advertising on Facebook is aimed at signaling candidate viability.

The growth of digital media in political advertising has broader implications for the ways in which candidates signal viability and build momentum. Digital advertising has proved highly effective for small-dollar fundraising, with television advertising most effective for getting out the vote. During the 2008 primary season, Barack Obama exceeded fundraising expectations by raising nearly \$30 million 
%CUT online 
from individuals donating \$100 or less ~\parencite{luo2008a}. 
%CUT More recently, 
Donald Trump's 2020 re-election campaign has fully embraced this strategy, spending heavily on digital advertising and receiving 725,000 small-dollar donations
%CUT averaging \$48 each 
~\parencite{karni2019a}. While prior research has focused on the total amount of money raised by candidates as a sign of viability, the recent increase in digital advertising may signal that the quantity of unique donors has become a critical metric, particularly for less well-known candidates. 

Digital advertising may improve democratic governance by increasing candidates' access to primary contests. During the 2020 primary cycle, an unprecedented number of Democratic candidates utilized digital media in addition to other strategies to remain competitive.
%CUT for 6 to 14 months. 
Since candidates can reach large numbers of voters with relatively small budgets, future primary cycles may be characterized by a competitive field that takes much longer to narrow than in the past. Further, candidates who lose the nomination may still exert influence by participating in debates and shaping other candidates' platforms. 

On the other hand, digital platforms are believed by some to disrupt the democratic process and drive political polarization. 
%To maximize the number of ads shown and clicked, 
Facebook aims to display content they believe to be relevant, at the expense of apparently less germane material. Showing ads that the users do not already agree with might cause cognitive dissonance, drive the users off the platform and result in ``antigrowth''~\parencite{horwitz2020}. This approach
%might work well for advertising consumer products but 
--- when applied in the political context --- can lead to negative societal effects. People who are algorithmically prevented from seeing a different point of view will eventually have a reduced pool of information when making their choices at the polling station.  

To better understand the role of online advertising, we need a more comprehensive picture of how digital platforms are operated and governed, but also how they are used both by those who run the ads and those who consume them. 
%This work brings us another step closer.

% \bibliographystyle{apalike}
% \bibliography{sample}

\printbibliography

\newpage

\title{Evolution of Digital Advertising Strategies\\during the 2020 US Presidential Primary
\large{Supplementary Information}}

% The custom \author command takes THREE arguments:
% #1 = Author name
% #2 = Affiliation name
% #3 = Brief author profile, or anything that you'd usually put in a \thanks. Leave blank {} if there's nothing to be said.
\author{NaLette Brodnax$^1$ \and Piotr Sapiezynski$^2$}
\date{\small{$^1$McCourt School of Public Policy, Georgetown University \\
      $^2$Khoury College of Computer Sciences, Northeastern University}}
      
\maketitle

\setcounter{section}{0}
\renewcommand{\thesection}{}
%\section*{Supplementary Information}

\setcounter{table}{0}
\renewcommand{\thetable}{S\arabic{table}}

\setcounter{figure}{0}
\renewcommand{\thefigure}{S\arabic{figure}}

\setcounter{section}{0}
\renewcommand{\thesection}{S\arabic{section}}
\begin{refsection}
In Section~\ref{sec:si_advertising} we provide additional information about the life-cycle of advertisements on Facebook, from creation to the mechanics of online ad auctions.
In Section~\ref{sec:si_dataset_fb} we describe the data we collected from the Facebook Ad Archive as well as additional information about candidates that took part in the 2020 US presidential primaries.
Section~\ref{sec:si_measures} details the construction of the dependent variable and its relation to established metrics such as Gross Rating Points (GRP).
Finally, in Section~\ref{sec:si_results} we compare our findings to the strategies observed in television advertising during previous primary cycles.

\section{Advertising on Facebook}\label{sec:si_advertising}
Social media platforms have taken different approaches to political advertising. Facebook's official stance is that political advertising is an important form of free speech that should be permitted even if it contains false claims~\parencite{schiffer2019}.
Twitter took the opposite stance---that paid speech is not free speech---and outright banned all political advertising ~\parencite{conger2019a}.
Political campaigns can still publish content on Twitter but cannot pay to increase its reach.
Google runs political ads, but only recently began publishing data as part of its own transparency project. This study utilizes data from Facebook, which reports all political advertisements published on its platform going back to May 2018. 

One of the major differences between online ad platforms and the traditional media is the precision with which the advertiser can describe their ideal audience. Beyond basic demographics such as age, gender, and location, one can also target by inferred interests, including but not limited to political leaning.
During the ad creation process, advertisers can use several targeting options at once to refine the audience; doing so is often referred to as {\em micro-targeting}.
For example, a political campaign might upload the list of people who signed up for their newsletter, ask Facebook to find users who ``look alike'' (but have not yet signed up), and further refine that audience to target the middle-age and older populations who are historically more likely to vote.
The campaign might also use publicly available voter registration data to create a custom audience consisting of registered voters of the opposing party, and then discourage those people from voting.
Campaigns' real-world use of these features has been discussed in the popular press~\parencite{martinez2018,lapowsky2018} but is understudied academically. 

There are two distinct steps in the life-cycle of an ad on Facebook: creation and delivery.

\subsection{Ad creation}
During ad creation, the advertiser makes a series of decisions that influence delivery, including:
\subsubsection{Ad creative} The advertiser designs the appearance of the ad: its text, media (image or video), and the link that the users will open upon clicking the ad.

\subsubsection{Ad placement} There are multiple avenues through which Facebook shows ads to users, and the advertiser can choose any number of them among: desktop News Feed, mobile News Feed, Messenger, Marketplace, Instagram, online ads outside of the Facebook family of products, etc. 

\subsubsection{Targeting} A tool called {\em Custom Audiences} allows advertisers to list the target users directly by their personally identifiable information such as their email address, phone number, or a combination of name, age, and zip code.
Advertisers who embed Facebook tracking technology on their websites can also create Custom Audiences consisting of Facebook users who visited the advertiser's website.
Finally, an advertiser can ask Facebook to extend their source Custom Audience into a {\em Lookalike audience} consisting of users who Facebook deems similar to those in the source custom audience.
The exact mechanism of what constitutes ``similarity'' is proprietary but likely considers the social network structure, Facebook activity, browsing history, and demographic information.

\subsubsection{Budgeting} The advertiser decides how much they are willing to spend on an ad (in total or daily) and whether they want the budget to be distributed evenly over time, or spent as quickly as possible.

\subsubsection{Optimization goal} The advertiser declares the goal of the campaign: to raise awareness (by showing the ad to as many users as possible), drive traffic (showing the ad to people who are more likely to click on it), or encourage even more precise action, like purchasing a product, or making a donation.

\subsection{Ad delivery}
The Facebook Library contains all political ads starting from May 2018 in the US and select other countries.
Each ad is described by its creative elements (text, link, and image or video), as well as rough delivery statistics:
\begin{enumerate}
	\item start time;
	\item estimate of spend (in ranges of USD 0-100, 100-500, 500-1000, 1000-5000, 5000-10000, 10000-50000, 50000+);
	\item estimate of impressions (in ranges of 0-1000, 1000-5000, 5000-10000, ...);
	\item gender and age breakdown (fraction of impressions by women, men, and users with unknown gender, further broken down by age in ranges 18-24, 25-34, 35-44, 45-54, 55-64, 65+);
	\item geographical breakdown (fraction of impressions per state);
\end{enumerate}

Facebook runs live auctions to determine which ads  users see. 
While these auctions were traditionally based on pricing, Facebook now considers many additional factors, among them, the inferred relevance of the ad to a particular user.
As a result, ads deemed relevant to users might win the auctions with lower bids, whereas apparently irrelevant ads might be financially penalized. %\footnote{See  Facebook documentation for more details: https://www.facebook.com/business/news/relevance-score}
The platforms might also consider the expected network value of the content---how likely it is to be shared.
On the surface, auctions encourage advertisers to create more relevant content, potentially creating a better online experience for the users.
On the other hand, this system can produce discriminatory skews when applied to life-opportunity ads such as education~\parencite{lambrecht-2019-algorithmic}, employment, or housing~\parencite{ali-2019-discrimination}.
Political ads can be presented in a skewed manner due to delivery optimization as well~\parencite{ali-2019-delivery}. An ad with the same content and targeting parameters will deliver to different demographics depending on how much the advertiser is willing to spend~\parencite{ali-2019-discrimination}.
Ali et al. show that it is as much as three times more expensive to show ads about Bernie Sanders to a conservative audience than it is to show the same audience ads about Donald Trump.

\section{Data: Facebook advertising}\label{sec:si_dataset_fb}

\subsection{Facebook Ad Library}
\begin{figure}[ht]
\centering
\includegraphics[width=\textwidth]{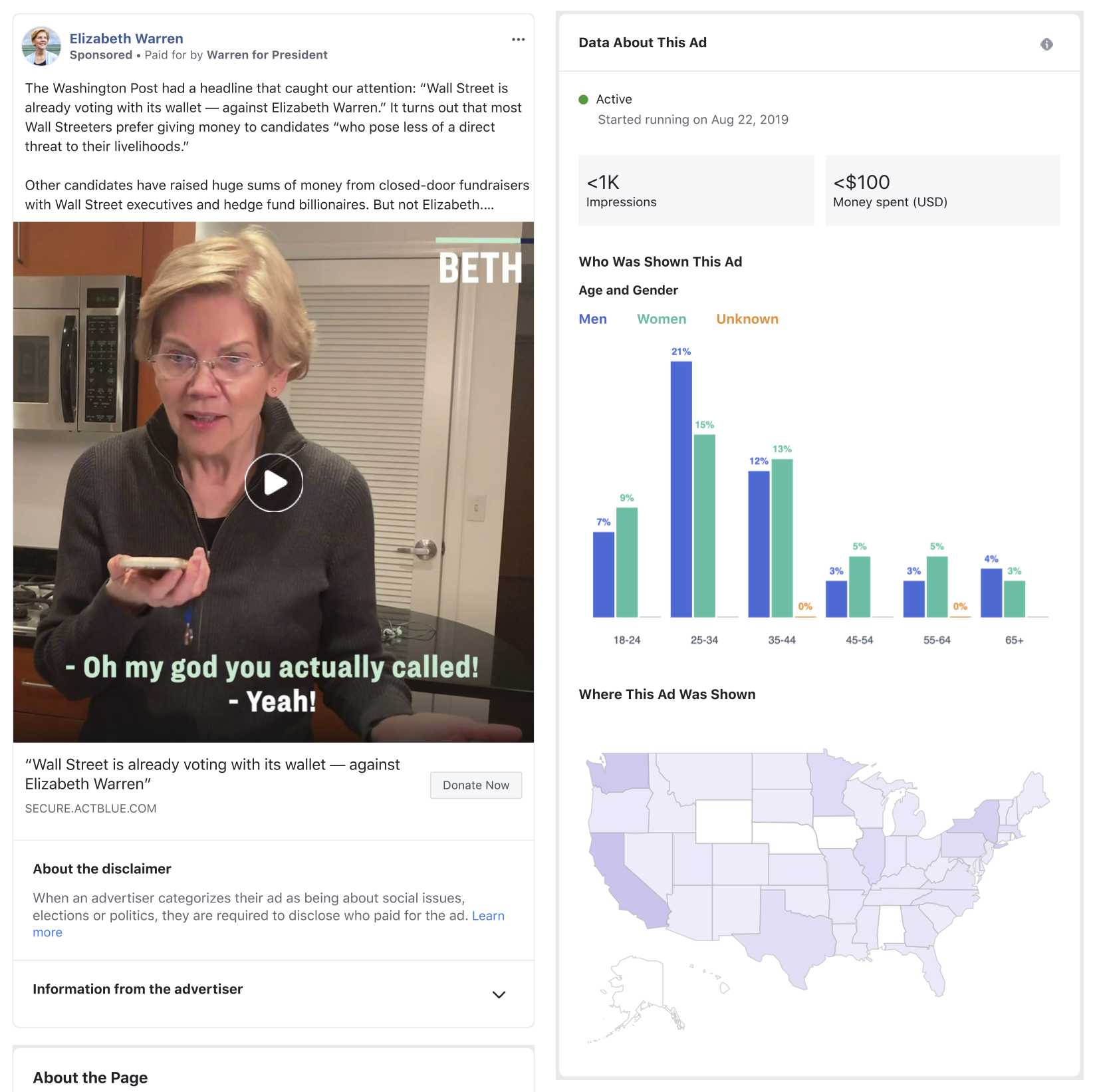}
\caption{Sample web interface from the Facebook Advertising Library}
\label{fig:archive}
\end{figure}

We use the Facebook Ad Library, the platform's official Application Programming Interface (API), to programmatically obtain ads published by 26 official presidential campaign accounts. Data for each ad include its creative elements (text, link, and image or video) as well as rough delivery statistics: start time, estimated spend, estimated impressions, gender and age breakdown, and geographical breakdown. See Fig.~\ref{fig:archive} and detailed descriptions of the API parameters and limitations. 
In total, our dataset contains 571,705 ads.

\subsubsection{Limitations}
The Facebook Ad Library provides information about the location and demographics of the audience that ultimately saw each ad.
Unless the advertiser specifies a budget that allows them to reach the entire targeted audience, each ad is shown to a subset. 
As we explain above, this subset is not selected strictly randomly.
Instead, Facebook attempts to preferentially show the ads to those users who Facebook deems relevant.
As a result, it is not possible to directly observe how much of the differences in geographical distribution of the ads are due to a campaign's deliberate strategy.
The delivery statistics are, instead, a proxy for campaign targeting.
Still, based on the information that \textit{is} available in the Ad Library we find strong evidence that the advertisers do target users by location and that this targeting is the leading factor in delivery.
First, if the effect was fully attributable to content-based optimization, we would see that similar ads deliver to similar demographics; but that is not the case. Upon closer inspection of the Ad Library, we see identical ads, each delivering to audiences located in different states. 
Second, we note that that 45\% of ads in our dataset were each only delivered in one state.
It is improbable that Facebook's delivery algorithm would deem an ad exclusively relevant to people in a particular state without clear direction from the advertiser.
Therefore, we conclude that deliberate geographic targeting by campaigns is a driving factor in the effects that we observe.

\subsection{Candidate characteristics}
\begin{figure}[ht]
\centering
\includegraphics[width=0.6\textwidth]{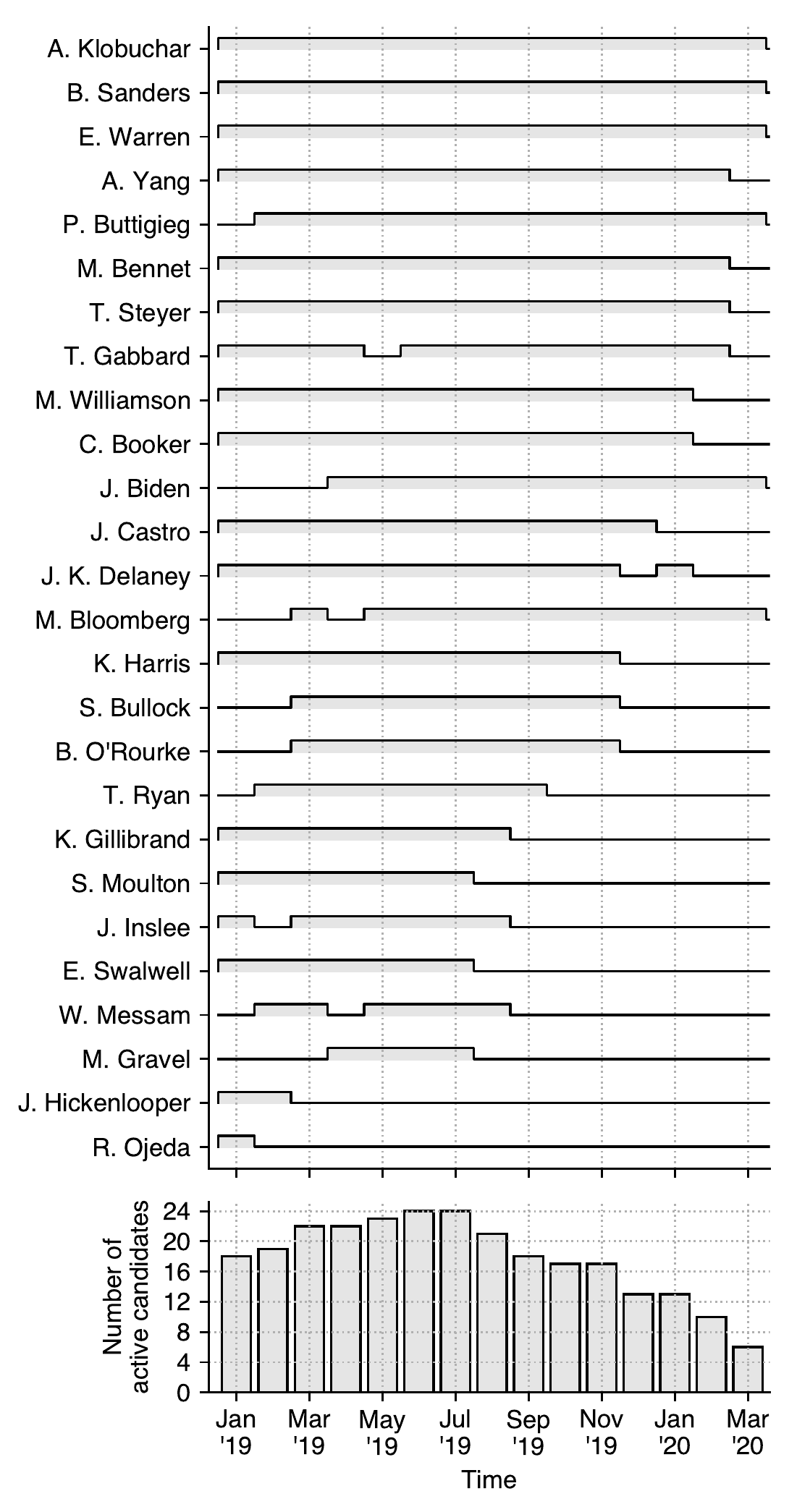}
\caption{The dataset contains information about 26 Democratic candidates, most of whom did not actively advertise for the entire period of observation (Jan 2019 - March 2020)}
\label{fig:candidate_count}
\end{figure}

A historically large number of major candidates participated in the 2020 Democratic primary: 29, in contrast to five in 2016 and eight in 2008 ~\parencite{fec2020a}.\footnote{No Democratic candidates ran against incumbent Barack Obama in the 2012 primary.} 
The primary campaign cycle began in January 2019 with 18 candidates and gradually increased to a mid-year peak of 24, before declining steadily to just 6 candidates publishing ads leading up to Super Tuesday in March 2020. 
Since many candidates were already elected officials, we exclude ads published prior to January 2019 to distinguish between the 2020 primary season and the 2018 midterm election season. 
We consider major candidates to be those that reported at least \$100,000 in contributions from individuals other than the candidate. 
Fig.~\ref{fig:candidate_count} shows the monthly activity of candidates competing for the nomination and running Facebook ads. As indicated by the raised gray blocks, most purchased ads over a period of six months or longer between January 2019 and March 2020. Just three candidates---Amy Klobuchar, Bernie Sanders, and Elizabeth Warren---purchased ads in each of the months included in our study. Roughly half of candidates spent consistently across 2019, while the remaining gradually tapered off in the third and fourth quarters of 2019. 
% The lower panel of Fig.~~\ref{fig:candidate_count} shows the total number of candidates actively purchasing ads in the given month. 

\begin{figure}[ht]
\centering
\includegraphics[width=0.6\textwidth]{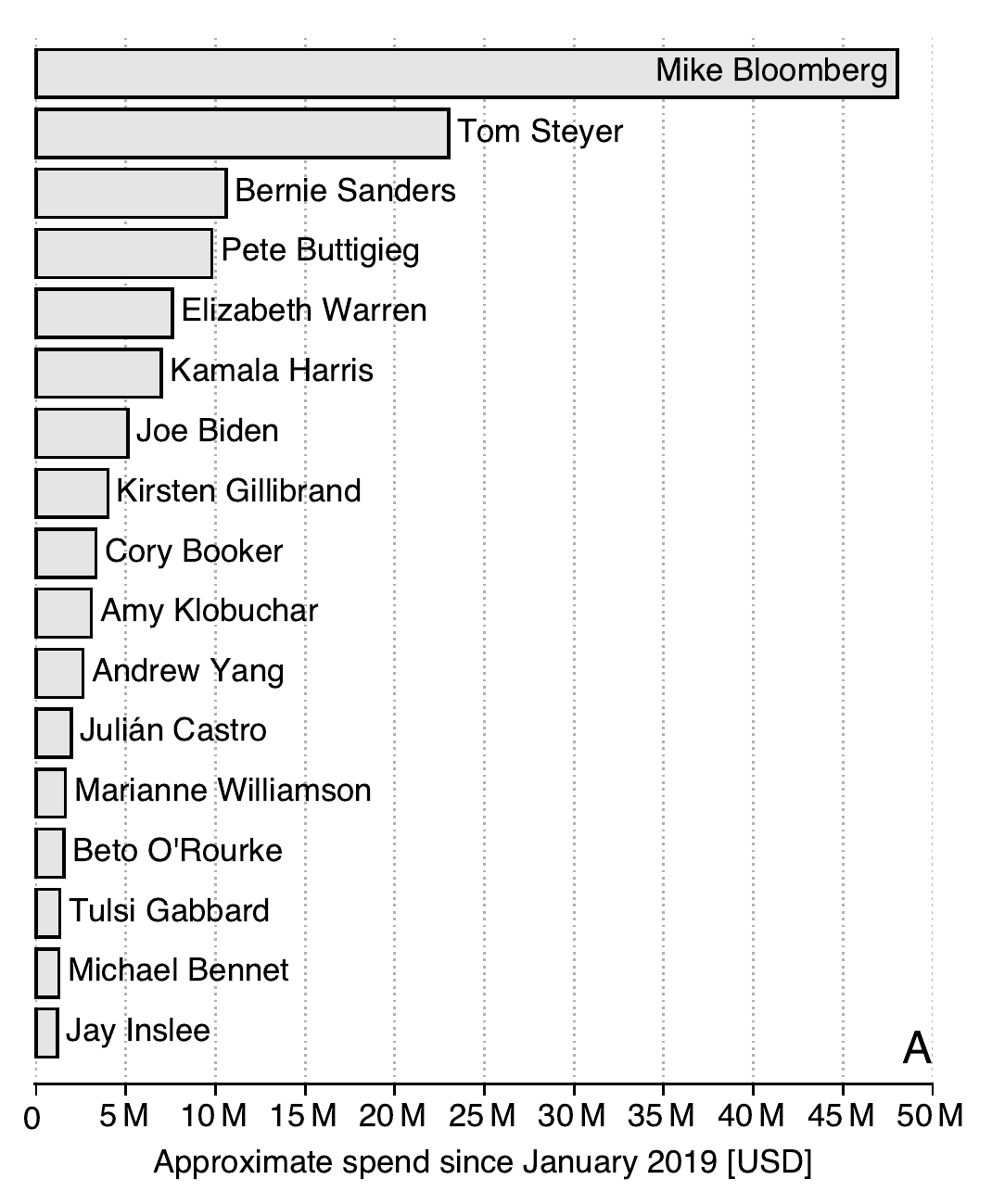}
\includegraphics[width=0.6\textwidth]{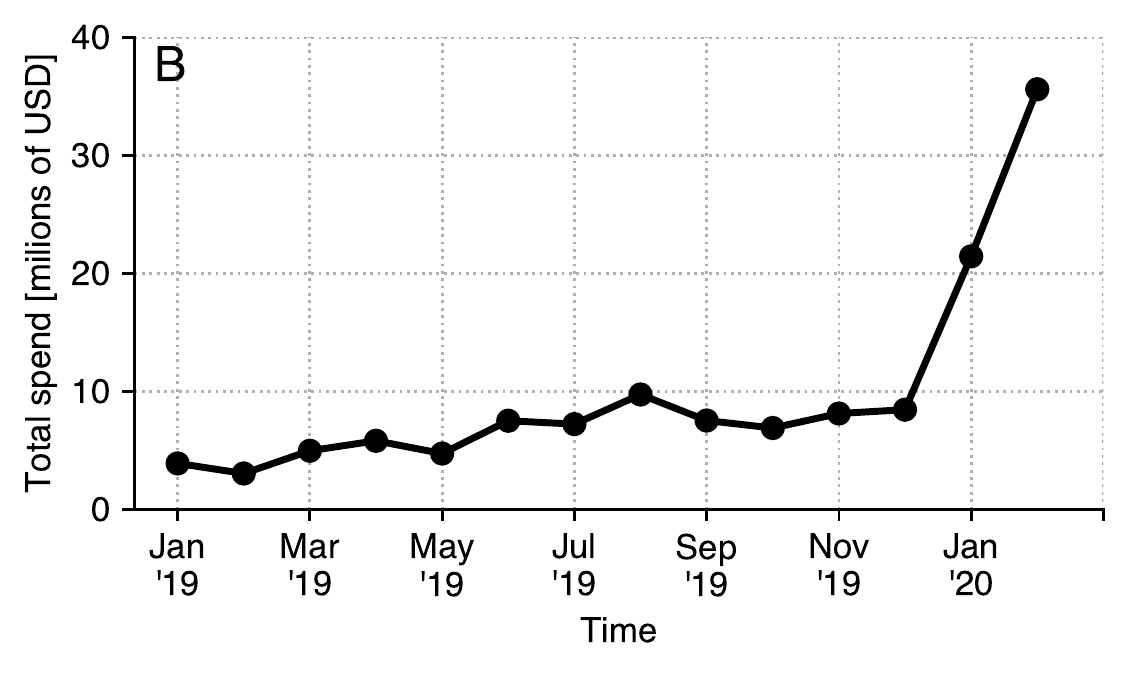}
\caption{Approximate Facebook ad expenditure among campaigns spending at least \$1 million (A) and total spending per month for all campaigns (B) since January 2019.}
\label{fig:total_budgets}
\end{figure}

Fig.~\ref{fig:total_budgets} shows estimated spending on Facebook advertising for campaigns with budgets of at least \$1 million, as well as the total monthly spending for all candidates over time. Ad spending varied widely across campaigns, with the vast majority spending \$5 million or less. While spending gradually increased over the course of the primary cycle, the large spikes in early 2020 can be attributed to wealthy candidates such as Michael Bloomberg. Notably, the presumptive Democratic nominee, Joe Biden, ranked 7th among the highest spending candidates, suggesting that digital ad spending alone may not be a strong driver of primary success.  

\subsection{State characteristics}

Our analysis examines the extent to which candidates target voters in states with characteristics that correspond to our hypotheses. Table~\ref{table:states} provides a breakdown of candidate home states, swing states, and states with early primaries. We consider a candidate's home to be the state in which he or she held an elected position or resided as of January 1, 2019, and we consider swing states to be those identified by poll aggregator FiveThirtyEight during the 2016 general election cycle ~\parencite{silver2016a}.

\begin{table}[!htbp] \centering 
  \caption{State Characteristics} 
  \label{table:states} 
\begin{tabular}{p{4.5cm} p{6cm}}
\toprule
 Candidate home state & AK, CA, CO, DE, FL, HI, IL, IN, MD, MA, MN, NJ, NY, OH, TX, VT, WA, WV \\\hline\\[-1.8ex] 
 Swing state in 2016 & AZ, CO, FL, GA, IA, ME, MI, MN, NV, NH, NC, OH, PA, UT, VA, WI\\ \hline\\[-1.8ex] 
 February primary & IA, NH, NV, SC \\ \hline\\[-1.8ex] 
 Super Tuesday primary & AL, AR, CA, CO, GA, MA, MN, NC, OK, TN, TX, UT, VT, VA \\
 \bottomrule
 \end{tabular}
\end{table} 

Of the four states that hold caucuses or primaries in February, Iowa and New Hampshire are considered key litmus tests for candidate viability, even though the primary delegates account for just 3.3\% of the 1,991 delegates needed to secure the Democratic nomination. South Carolina's primary at the end of February is also important to candidates because it is the first primary with a large block of black voters.  On Super Tuesday, 14 states hold primaries to allocate 1,600 delegates---the largest number on any single day of the season.
Following Super Tuesday 2020 only two candidates remained: presumptive nominee Joe Biden and Bernie Sanders, who dropped out a month later.

\section{Measures: Dependent Variable Construction}
\label{sec:si_measures}
To construct our dependent variable, we draw on the approach of television advertisers. Television media buyers use a metric called gross ratings points (GRPs) to measure the reach of an individual ad ~\parencite{fowler2018a}. An ad receives one GRP for each percentage point of the target audience reached by the ad's airing. These GRPs can then be aggregated to express an advertiser's total budget. We extend this approach to construct an aggregated measure of digital reach but with two changes. First, we express the campaign's budget as a proportion. Let $Y_{is}$ be the population-normalized proportion of the advertising budget allocated to state $s$ by campaign $i$. If a campaign spends money in direct proportion to the number of inhabitants, then $Y_{is}=1$. When campaigns spend more or less than expected given a state's population, then $Y_{is}\neq 1$
We use the budget proportion as our dependent variable rather than the dollar amount because it allows us to compare candidates whose budgets vary by orders of magnitude.
We normalize the fraction by each state's population to compare how much value campaigns assign to voters living in states whose populations vary by orders of magnitude.
As pointed out by Gimpel et al.~\parencite{gimpel2006a}, the total amount of donations that come from a state is highly correlated with that state's population regardless of the campaign's leaning or strategy. 
Similarly, in our data the correlation between per-state budgets and the state populations across all campaigns is on average $\rho=0.61$. 
Normalizing by population allows us to take a more nuanced view on the geographic aspects of the strategic ad buys.

\section{Results: Comparison to Prior Elections}
\label{sec:si_results}
\begin{figure}
\centering
\includegraphics[width=0.6\textwidth]{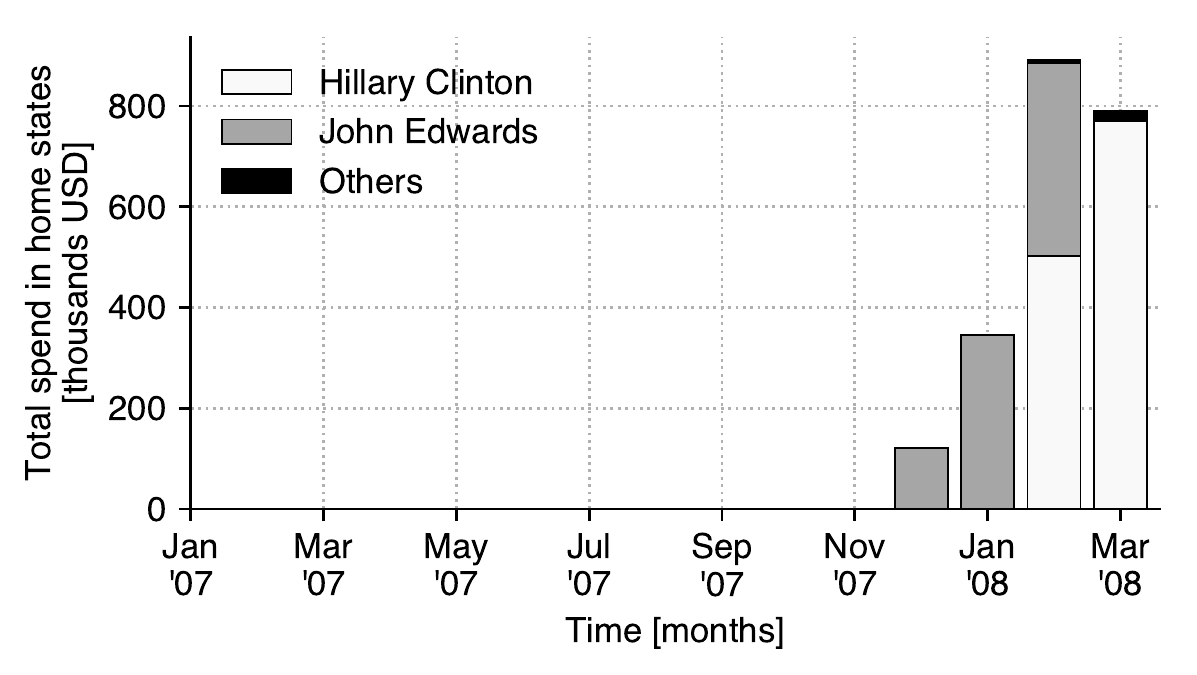}
\caption{During the 2008 presidential primary cycle, just four candidates purchased television airtime in their home states, all in the weeks leading up to the first primaries and caucuses in February. The absence of early-cycle spending on TV ads suggests that campaigns are employing a new strategy of targeting candidates' home states to raise money and awareness.}
\label{fig:tv_spend}
\end{figure}

During the 2008 primary cycle, just 4 of the 13 major campaigns advertised in television markets where voters from the candidate's home state comprised the core audience. However, virtually no home state targeting occurred early in the primary cycle. 
As shown in Fig.~\ref{fig:tv_spend}, nearly all of these expenditures occurred towards the end of the 14-month period that ends with the February primaries. Hillary Clinton spent a total of \$35.5M on television advertisements during the primary. Of this, she spent \$1.3M (3.6\%) in her home state of New York in early 2008, 3.7\% of her budget. John Edwards spent \$850K in North Carolina in late 2007 and early 2008---10.3\% of his total budget of \$9.1M. Candidates Mike Huckabee of Arizona and Ron Paul of Texas spent less than half of one percent of their respective budgets of approximately \$3 million. Notably, Barack Obama, who won the nomination, spent more on television advertising than any other candidate (\$51.3M) but did not allocate any of this budget to target voters in his home state of Illinois. During the 2012 cycle, just one major candidate---Mitt Romney---advertised in his home state of Massachusetts. Romney spent \$636K, 5.48\% of his total advertising budget of \$11.6 million, all in early January 2012, over a year into the primary season. 

Comparisons between digital and television spending face important limitations. First, broadcast television markets are organized into geographic units called designated market areas (DMAs). DMAs can span multiple states, making it impossible to perfectly capture state-level spending. Further, although campaigns spend the vast majority of their advertising dollars with local broadcast stations, they may also purchase airtime from national or local cable networks, which utilize different geographic units. Second, campaigns often purchase airtime several months in advance in order to secure the lowest rates and avoid being locked out of prime airtime by competing advertisers ~\parencite{fowler2018a}. This may explain why television spending is more concentrated in the weeks leading up to the first caucuses and primaries, as campaigns prioritize encouraging voters to turn out on election day. Even with the limited utility of comparing digital and television media spending, the absence of early-cycle TV advertising suggests that digital media has ushered in a set of new and lower cost opportunities for campaigns to expand their reach early on and throughout the primary cycle.

%%% Each figure should be on its own page

\begin{figure}
\centering
\includegraphics[width=0.6\textwidth]{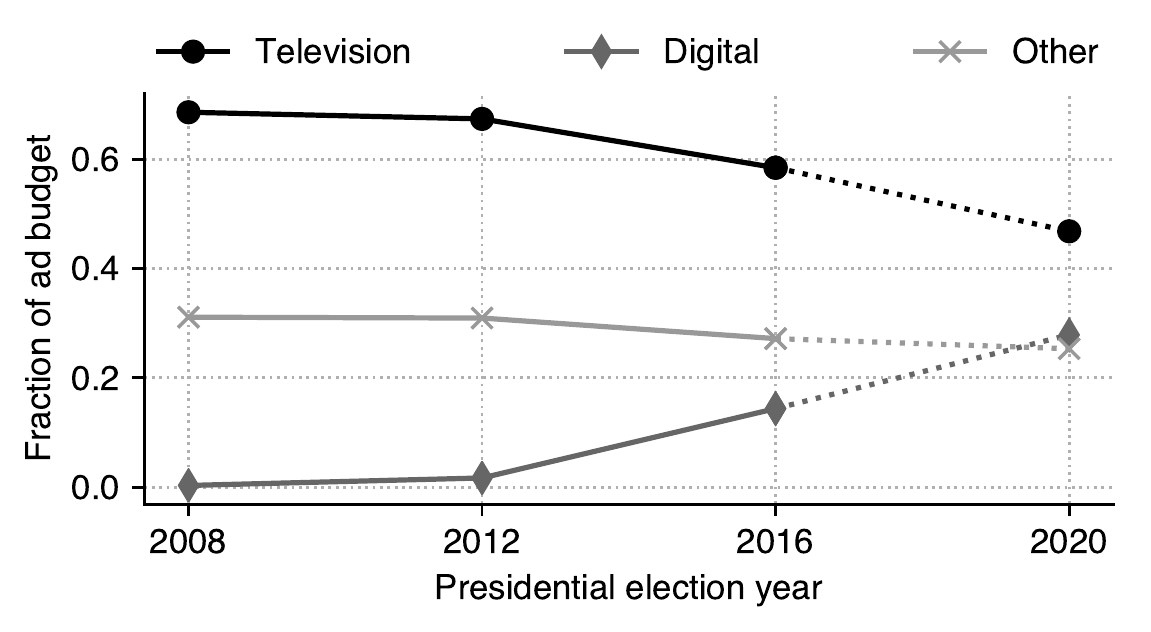}
\caption{Online advertising has become more central to political advertising budgets, from below 1\% in 2008 to 14.4\% in 2016. It is projected to reach 28\% in 2020 and surpass the combined spending on political ads in press, radio, direct mail, telemarketing and others combined~\parencite{cassino2017a}.}
	\label{fig:budget_fracs}
\end{figure}

\printbibliography

@book{la-raja2015a,
	Author = {{La Raja}, Raymond J and Schaffner, Brian F},
	Publisher = {University of Michigan Press},
	Title = {{Campaign Finance and Political Polarization: When Purists Prevail}},
	Year = {2015}}

@article{dowdle2009a,
	Author = {Dowdle, Andrew J and Adkins, Randall E and Steger, Wayne P},
	Journal = {Political Research Quarterly},
	Number = {1},
	Pages = {77--91},
	Title = {{The Viability Primary: Modeling Candidate Support Before the Primaries}},
	Volume = {62},
	Year = {2009}}

@book{bartels1988a,
	Author = {Bartels, Larry M},
	Publisher = {Princeton University Press},
	Title = {{Presidential Primaries and the Dynamics of Public Choice}},
	Year = {1988}}

@article{lin2017a,
	Author = {Lin, Yu-Ru and Kennedy, Ryan and Lazer, David},
	Journal = {Research and Politics},
	Pages = {1--8},
	Title = {{The geography of money and politics: Population density, social networks, and political contributions}},
	Year = {2017}}

@article{cho2007a,
	Author = {Cho, Wendy K Tam and Gimpel, James G},
	Journal = {American Journal of Political Science},
	Number = {2},
	Pages = {255--268},
	Title = {{Prospecting for (Campaign) Gold}},
	Volume = {51},
	Year = {2007}}

@article{gimpel2006a,
	Author = {Gimpel, James G and Lee, Frances E and Kaminski, Joshua},
	Journal = {Journal of Politics},
	Number = {3},
	Pages = {626--639},
	Title = {{The Political Geography of Campaign Contributions in American Politics}},
	Volume = {68},
	Year = {2006}}

@misc{karni2019a,
	Author = {Karni, Annie and Haberman, Maggie},
	Howpublished = {https://www.nytimes.com/2019/07/02/us/politics/trump-fundraising.html},
	Month = {July},
	Title = {{Trump and R.N.C. Raised \$105 Million in 2nd Quarter, a Sign He Will Have Far More Money Than in 2016}},
	Year = {2019}}

@misc{luo2008a,
	Author = {Luo, Michael},
	Howpublished = {https://www.nytimes.com/2008/02/20/us/politics/20obama.html},
	Month = {February},
	Title = {{Small Online Contributions Add Up to Huge Fund-Raising Edge for Obama}},
	Year = {2008}}

@misc{horwitz2020,
	Author = {Horwitz, Jeff and Seetharaman, Deepa},
	Howpublished = {https://www.wsj.com/articles/facebook-knows-it-encourages-division-top-executives-nixed-solutions-11590507499?mod=djemalertNEWS},
	Month = {May},
	Title = {{Facebook Executives Shut Down Efforts to Make the Site Less Divisive }},
	Year = {2020}}

@book{fowler2018a,
	Author = {Fowler, Erika Franklin and Franz, Michael M and Ridout, Travis N},
	Publisher = {New York : Routledge},
	Title = {{Political Advertising in the United States}},
	Year = {2018}}

@misc{dnc2019a,
	Author = {{Democratic National Committee}},
	Howpublished = {https://democrats.org/news/dnc-announces-details-for-the-first-two-presidential-primary-debates/},
	Month = {February},
	Title = {{DNC Announces Details for the First Two Presidential Primary Debates}},
	Year = {2019}}

@misc{fowler2017a,
	Author = {Fowler, Erika Franklin and Franz, Michael M and Ridout, Travis N},
	Howpublished = {{Version 1.0 [dataset] Middletown, CT: The Wesleyan Media Project, Department of Government at Wesleyan University}},
	Title = {{Presidential Political Advertising in the 2012 cycle}},
	Year = {2017}}

@misc{fec2020a,
	Author = {{US Federal Election Commission}},
	Howpublished = {https://www.fec.gov/data/browse-data/?tab=candidates},
	Month = {March},
	Title = {{Presidential Map Candidates: Overall Candidate Summary}},
	Year = {2020}}

@misc{silver2016a,
	Author = {Silver, Nate},
	Howpublished = {https://fivethirtyeight.com/features/the-odds-of-an-electoral-college-popular-vote-split-are-increasing/},
	Month = {October},
	Title = {{The Odds of an Electoral College-Popular Vote Split Are Increasing}},
	Year = {2016}}

@misc{conger2019a,
	Author = {Conger, Kate},
	Howpublished = {https://www.nytimes.com/2019/10/30/technology/twitter-political-ads-ban.html},
	Month = {October},
	Title = {{Twitter Will Ban All Political Ads, C.E.O. Jack Dorsey Says}},
	Year = {2019}}

@misc{almukhtar2019a,
	Author = {Almukhtar, Sarah and Martin, Jonathan and Stevens, Matt},
	Howpublished = {https://www.nytimes.com/interactive/2019/us/elections/2020-presidential-election-calendar.html?rref=collection\%2Fnewseventcollection\%2F2020-election},
	Month = {February},
	Title = {2020 Presidential Primary Election Calendar},
	Year = {2019}}

@misc{scherer2019a,
	Author = {Scherer, Michael},
	Howpublished = {https://www.washingtonpost.com/politics/september-debate-rules-could-winnow-2020-democratic-field/2019/05/28/93601076-81a5-11e9-95a9-e2c830afe24f\_story.html},
	Month = {May},
	Title = {Tougher new debate rules could dramatically winnow Democratic presidential field},
	Year = {2019}}

@techreport{cassino2017a,
	Author = {Cassino, Kip},
	Institution = {Borrell Associates},
	Title = {{The Final Analysis: What Happened to Political Advertising in 2016 (And Forever)}},
	Year = {2017}}

@misc{goldstein2011a,
	Author = {Goldstein, Kenneth and Niebler, Sarah and Neilheisel, Jacob and Holleque, Matthew},
	Howpublished = {{Combined File [dataset]. Initial release. Madison, WI: The University of Wisconsin Advertising Project, the Department of Political Science at the University of Wisconsin-Madison}},
	Title = {{Presidential, Congressional, and Gubernatorial Advertising, 2008}},
	Year = {2011}}

@misc{erdody2018a,
	Author = {Erdody, Lindsey},
	Howpublished = {https://www.ibj.com/articles/70545-political-campaigns-boost-investment-in-social-media-ads},
	Journal = {Indianapolis Business Journal},
	Month = {September},
	Title = {Political campaigns boost investment in social media ads},
	Year = {2018}}

@article{shaw1999a,
	Author = {Shaw, Daron R},
	Journal = {Journal of Politics},
	Number = {4},
	Pages = {893--913},
	Title = {The Methods Behind the Madness: Presidential Electoral College Strategies, 1988-1996},
	Volume = {61},
	Year = {1999}}

@misc{schiffer2019,
	Author = {Schiffer, Zoe},
	Howpublished = {https://www.theverge.com/2019/10/17/20919223/mark-zuckerberg-facebook-speech-live-politics-threats-free-expression},
	Month = {October},
	Title = {{Mark Zuckerberg on lies in political ads: `I don't think it's right for a private company to censor politicians'}},
	Year = {2019}}

@misc{martinez2018,
	Author = {Mart{\'{i}}nez, Antonio Garc{\'{i}}a},
	Howpublished = {https://www.wired.com/story/how-trump-conquered-facebookwithout-russian-ads/},
	Month = {February},
	Title = {{How Trump Conquered Facebook - Without Russian Ads}},
	Year = {2018}}

@misc{lapowsky2018,
	Author = {Lapowsky, Issie},
	Howpublished = {https://www.wired.com/story/facebook-trump-clinton-campaign-ad-cpms/},
	Month = {February},
	Title = {What Facebook Isn't Saying About Trump and Clinton's Campaign Ads},
	Year = {2018}}

@article{aldrich2009a,
	Author = {Aldrich, John},
	Journal = {PS: Political Science and Politics},
	Number = {1},
	Pages = {33--38},
	Title = {The Invisible Primary and Its Effects on Democratic Choice},
	Volume = {42},
	Year = {2009}}

@book{shaw2006a,
	Author = {Shaw, Daron R},
	Publisher = {University of Chicago Press},
	Title = {The Race to 270: The Electoral College and the Campaign Strategies of 2000 and 2004},
	Year = {2006}}

@article{urban2014a,
	Author = {Urban, Carly and Niebler, Sarah},
	Journal = {American Journal of Political Science},
	Number = {2},
	Pages = {322--336},
	Title = {Dollars on the Sidewalk: Should {U.S.} Presidential Candidates Advertise in Uncontested States?},
	Volume = {58},
	Year = {2014}}

@article{adkins2001a,
	Author = {Adkins, Randall E and Dowdle, Andrew J},
	Journal = {Political Research Quarterly},
	Number = {2},
	Pages = {431--444},
	Title = {How Important Are Iowa and New Hampshire to Winning Post-Reform Presidential Nominations?},
	Volume = {54},
	Year = {2001}}

@article{paolino2003a,
	Author = {Paolino, Philip and Shaw, Daron R},
	Journal = {PS: Political Science and Politics},
	Number = {2},
	Pages = {193--197},
	Title = {Can the Internet Help Outsider Candidates Win the Presidential Nomination?},
	Volume = {36},
	Year = {2003}}

@article{christenson2014a,
	Author = {Christenson, Dino P and Smidt, Corwin D and Panagopoulos, Costas},
	Journal = {Political Research Quarterly},
	Number = {1},
	Pages = {108--122},
	Title = {Deus ex Machina: Candidate Web Presence and the Presidential Nomination Campaign},
	Volume = {67},
	Year = {2014}}

@misc{ali-2019-discrimination,
	Author = {Muhammad Ali and Piotr Sapiezynski and Miranda Bogen and Aleksandra Korolova and Alan Mislove and Aaron Rieke},
	Month = {April},
	Note = {\url{https://arxiv.org/abs/1904.02095}},
	Title = {{Discrimination Through Optimization: How Facebook's Ad Delivery Can Lead To Skewed Outcomes}},
	Year = 2019}

@misc{ali-2019-delivery,
	Author = {Muhammad Ali and Piotr Sapiezynski and Aleksandra Korolova and Alan Mislove},
	Month = {November},
	Note = {\url{https://arxiv.org/abs/1904.02095}},
	Title = {{Ad Delivery Algorithms: The Hidden Arbiters of Political Messaging}},
	Year = 2019}

@misc{FacebookAdLibrary,
	Key = {Facebook Ad Library},
	Note = {\url{https://www.facebook.com/ads/library/}},
	Year = {2019},
	Title = {{Facebook Ad Library}}}

@article{lambrecht-2019-algorithmic,
	Author = {Lambrecht, Anja and Tucker, Catherine},
	Journal = {Management Science},
	Publisher = {INFORMS},
	Title = {Algorithmic Bias? An Empirical Study of Apparent Gender-Based Discrimination in the Display of STEM Career Ads},
	Year = {2019}}

@article{fowler2020,
	Author = {Fowler, Erika Franklin and Franz, Michael M and Martin, Gregory J and Peskowitz, Zachary and Ridout, Travis N},
	Journal = {American Political Science Review},
	Pages = {1--20},
	Title = {{Political Advertising Online and Offline}},
	Year = 2020}
\end{refsection}

\end{document}